\documentclass[letterpaper, 10 pt, conference]{ieeeconf}
\IEEEoverridecommandlockouts
\usepackage{algorithm}
\usepackage[noend]{algpseudocode}

\usepackage{booktabs} 
\usepackage{multirow} 
\usepackage{makecell} 
\usepackage{arydshln}
\usepackage{cancel}
\usepackage{ulem}

\newtheorem{theorem}{\textbf{Theorem}}

\newtheorem{remark}{\textbf{Remark}}

\renewcommand{\em}{\it}
\renewcommand{\emph}{\textit}

\usepackage{url}
\usepackage{amsmath,amssymb,amsfonts}
\usepackage{mathrsfs}
\usepackage{graphicx}
\usepackage{textcomp}
\usepackage{xcolor}
\def\BibTeX{{\rm B\kern-.05em{\sc i\kern-.025em b}\kern-.08em
    T\kern-.1667em\lower.7ex\hbox{E}\kern-.125emX}}
\usepackage{hyperref}       
\usepackage{url}            
\usepackage{booktabs}       
\usepackage{microtype}      %

\title{Orthogonal Nonnegative Matrix Factorization with Sparsity Constraints
}
\author{Salar Basiri$^{1}$, Alisina Bayati$^{1}$, Srinivasa M. Salapaka$^{1}$
\thanks{$^{1}$ University of Illinois at Urbana-Champaign, Urbana, Il, 61801, USA; Emails:
        {\small $\{$ sbasiri2, abayati2, salapaka$\}$@illinois.edu}. We acknowledge the support of National Aeronautics and
Space Administration under Grant NASA 80NSSC22M0070 for this work.}
}
\begin{document}
\maketitle

\begin{abstract}
This article presents a novel approach to solving the sparsity-constrained Orthogonal Nonnegative Matrix Factorization (SCONMF) problem, which requires decomposing a non-negative data matrix into the product of two lower-rank non-negative matrices,  $X=WH$, where the mixing matrix $H$  has orthogonal rows ($ HH^\top=I$), while also satisfying an upper bound on the number of nonzero elements in each row. By reformulating SCONMF as a capacity-constrained facility-location problem (CCFLP), the proposed method naturally integrates non-negativity, orthogonality, and sparsity constraints. Specifically, our approach integrates control-barrier function (CBF) based framework used for dynamic optimal control design problems with maximum-entropy-principle-based framework used for facility location problems to enforce these constraints while ensuring robust factorization.
Additionally, this work introduces a quantitative approach for determining the {\it true} rank of $W$ or $H$, equivalent to the number of {\em true} features—a critical aspect in ONMF applications where the number of features is unknown.
Simulations on various datasets demonstrate significantly improved factorizations with low reconstruction errors (as small as by 150 times) while strictly satisfying all constraints, outperforming existing methods that struggle with balancing accuracy and constraint adherence. 
\end{abstract}

\section{Introduction} \label{sec:Intro}

In various machine learning, data science, and signal processing applications involving large datasets, identifying common features across data members and quantifying their weights is critical. For instance, in compressing facial image collections, it is essential to identify typical facial features and the extent of their occurrence in each face.  Since data in fields like computer vision and bioinformatics is often nonnegative, Nonnegative Matrix Factorization (NMF) is a powerful tool for such tasks. NMF decomposes a nonnegative matrix into two lower-rank nonnegative matrices: one capturing key features and the other quantifying their contributions. The nonnegativity constraint enhances interpretability, making NMF especially effective for inherently nonnegative data types, including images \cite{Li_2019,8372956}, audio recordings \cite{makino2018audio,gloaguen2019road}, biomedical data \cite{Esposito_2021,Carmona_Saez_2006,Song_2021}, spectrometry representations \cite{Trindade_2017}, and other data types \cite{manu_nmf}. Unlike Singular Value Decomposition (SVD) and Principal Component Analysis (PCA), which allow negative values and rely on orthogonality, NMF produces part-based, interpretable representations. This makes it particularly useful for applications where negative components lack real-world meaning.

Given a large nonnegative data matrix $X \in \mathbb{R}^{d \times n}_+ $, NMF seeks to approximate it as the product of two low-rank nonnegative matrices $X \approx WH$, 
where $W \in \mathbb{R}^{d \times k}_+$ is the {\em feature matrix} and $H \in \mathbb{R}^{k \times n}_+$ is the {\em mixing matrix}. The rank $k$ (with $ k \ll \min(n,d)$) determines the number of features. Each column of $W$ represents a basis feature in the original data space, while each column of $H$ encodes the contribution of these features to the corresponding data point. Specifically, the $\ell^{\text{th}}$ column of $X$ is approximated as a weighted sum of features given by 
$\sum_{s=1}^k h_{s\ell} w_s.$ The quality of approximation is typically measured using the Frobenius norm $\|X-WH\|_F$.


Various additional constraints have been imposed on the matrices $W$ and $H$ in the literature. One important constraint is {\em orthogonality}, where in the single-orthogonality case, either $H$ (rows) or $W$ (columns) must be orthogonal. In the more restrictive double-orthogonality case, both matrices must have orthogonal rows and columns, respectively. Certain applications, such as those in signal processing and bioinformatics, specifically require orthogonal features \cite{carmona,Esposito_2021}.  
Another widely used constraint is {\em sparsity}, which refers to enforcing a large number of zero or near-zero elements in either the feature or mixing matrix. This constraint enhances interpretability and improves computational efficiency \cite{seperable}. While enforcing both orthogonality and nonnegativity naturally induces some level of sparsity, certain applications require a predefined minimum sparsity level while maintaining these constraints \cite{mixed}.

Various algorithms have been developed to solve the Orthogonal NMF (ONMF) problem, including methods based on iterative projection updates \cite{ding,HALS,pnmf}, gradient descent \cite{IONMF}, and multiplicative gradient projections \cite{nhl,onmfa}. Additionally, some approaches frame ONMF as a clustering problem \cite{app} or propose EM-like algorithms \cite{emonmf}.  
In the broader context of NMF, promoting sparsity has led to the incorporation of various penalty functions into the objective function as a regularization term. Most studies focus on $\ell_1, \ell_{\frac{1}{2}}$, or mixed $\ell_1/\ell_2$-norm constraints to enforce sparsity \cite{l1,l_5,mixed}, while comparatively less attention has been given to the $\ell_0$-"norm" measure \cite{l0}. The $\ell_0$-"norm" of a vector $z \in \mathbb{R}^n$ counts its nonzero elements, that is, $||x||_{\ell_0} = \sum_{i=1}^n \mathbb{I}(x_i \neq 0)$, where $\mathbb{I}$ is the indicator function. While not a proper norm due to its lack of homogeneity, it is widely used in the literature to quantify sparsity.


\par
To our knowledge, no existing work addresses the NMF problem while simultaneously enforcing both $\ell_0$-sparsity and orthogonality constraints. Current methods cannot impose distinct sparsity bounds for each feature individually. Moreover, only a few approaches ensure orthogonality, often at the cost of reconstruction accuracy or computational efficiency. Additionally, most existing methods are highly sensitive to the initialization of $W$ and $H$ and typically require to fix the number $k$ of features a priori.  However, computational time, reconstruction error, and their trade-offs are directly influenced by the choice of $k$, making it a crucial yet often restrictive parameter.

In this paper, we propose a mathematical framework for solving the ONMF problem while enforcing an $\ell_0$-"norm" sparsity constraint. Our approach is flexible and accommodates scenarios where the number of features is unknown or needs to be determined adaptively. 
Our key insight is to reinterpret sparsity-constrained ONMF (SCONMF) as a Capacity-Constrained Facility Location Problem (CCFLP) - a class of  $\mathcal{NP}$-hard Facility Location Problems.  In CCFLP,  each facility has a limited capacity that restricts the number or demand of consumers it can serve. The goal is to determine optimal facility locations and assignments to minimize overall costs, such as transportation distance or operational expenses, while ensuring that no facility exceeds its capacity. SCONMF is a CCFLP, where columns of $X$  represent consumer locations, columns of $W$ (feature vectors) represent facility locations; and the mixing matrix $H$ encodes the assignments. 

We leverage a Maximum-Entropy Principle (MEP) based Deterministic Annealing (DA) algorithm—widely used in data compression and pattern recognition \cite{rose}—to solve FLPs. DA uses an iterative annealing process where, at each iteration, every consumer is associated with all facilities via a probability distribution. This distribution, along with facility locations, is obtained by solving a relaxed optimization problem, using the previous iteration's solution as an improved initial guess. In initial iterations, high-entropy distributions ensure equitable associations across facilities, reducing sensitivity to initialization; at later iterations, as entropy decreases, the distributions harden until each consumer is assigned to a single facility.
However, standard DA does not handle the {\em inequality} constraints of capacity in CCFLPs. To overcome this, we reformulate the static relaxed CCFLP (and thus the SCONMF problem) at each iteration as a constrained dynamic control problem. We demonstrate that solutions to this dynamic problem satisfy the Karush-Kuhn-Tucker (KKT) conditions of the original CCFLP. To solve the constrained dynamic problem efficiently, we employ a Control Barrier Functions (CBF)-based framework \cite{8796030,7782377}. This adaptation allows us to compute facility locations and probability distributions at each iteration, thereby extending DA to handle sparsity-constrained ONMF problems.

We assert that posing the SCONMF as a CCFLP and solving it through MEP provides remarkable advantages, such as guaranteeing nonnegativity and orthogonality, while maintaining invariance to initialization. Furthermore, Our method evolves hierarchically, enabling determination of the {\em true} number of features. 
At initial iterations, all $k$ features are identical, but with the iterations,  distinct features emerge, and reconstruction error $\|X - WH\|_F^2$ decreases. In \cite{truenumber}, we showed (in the FLP context)  that beyond a certain $k^\star$, reconstruction error reduction for every additional distinct feature becomes relatively negligible. This {\em true} number $k^\star$ is determinable from our algorithm, and also corresponds to the number of distinct features that persists over the widest range of reconstruction error values \cite{truenumber}.


\par
In simulations on synthetic and standard datasets, our algorithm outperformed other ONMF methods by achieving state-of-the-art reconstruction error, full orthogonality, and higher sparsity—all while improving computational efficiency. For example, on synthetic data, our method achieved a reconstruction error over 150 times smaller than the average of other methods, exhibited complete orthogonality, attained the highest sparsity, and ran the fastest. Additionally, it delivered up to 175\% higher sparsity compared to competing approaches.
\section{Problem Formulation} \label{sec:problem formulation}
Denote $\mathcal{D}_+^k$ as the set of $k \times k$ diagonal matrices with positive diagonal entries, and $\mathcal{P}^k$ the set of $k \times k$ permutation matrices (square matrices with exactly one entry of 1 in each row \emph{and} column, and 0 elsewhere). Define the set of generalized permutation matrices $\Delta_+^k := \{DP \ | D \in \mathcal{D}^k_+, P \in \mathcal{P}^k\}$. Consider the data matrix $X \in \mathbb{R}^{d \times n}_+$ that we want to approximate by the product of two nonnegative matrices $W \in \mathbb{R}^{d \times k}_+$ and $H \in \mathbb{R}^{k \times n}_+$.
 The ONMF poses the following optimization problem:
\begin{align}
    \min_{W,H} \ &D(X,WH) \label{ONMF} 
    \ \textrm{s.t.} \ HH^{\top}=I_k; \  
     W_{ij},H_{ij} \geq 0
\end{align}
 where $I_k$ is $k\times k$ identity matrix, $D(X,WH)$ is the distance function (representing the reconstruction error), and is often chosen to be the squared Frobenius norm $||X-WH||_F^2$. The orthogonality constraint may be replaced by $W^{\top}W=I_k$.
 \remark
 Without the orthogonality constraint $HH^\top=I_k$ in (\ref{ONMF}), if a solution $(\bar{W},\bar{H})$ exists, there exists a large number of square matrices $B \in \mathcal{B}$ ($B$ needs to be square to preserve inner dimension $k$) such that any other pair of the form $(\bar{W}B^{-1},B\bar{H})$ is also a solution to the problem as long as nonnegativity of factors are preserved ($ \Delta_+^k \subset \mathcal{B}$).
 However, this degree of freedom is removed with the presence of the orthogonality constraint $HH^\top=I_k$, and $B$ is restricted to be in $\mathcal{P}^k$, making the solution unique up to permutation. For proof, see Proposition 1 in \cite{ding}.

For ONMF, the constraint \( HH^\top = I_k \) is often overly restrictive, as it is sufficient for the rows of \( H \) to be orthogonal without requiring unit length. Therefore, a relaxed formulation is commonly used, replacing \( HH^\top = I \) with \( HH^\top \in \mathcal{D}_+^{k} \), which typically results in lower reconstruction errors compared to the original ONMF formulation (\ref{ONMF}).
Moreover, we can introduce the sparsity constraints on $H$, requiring number of non-zero values of each row denoted by $c_j :=||H_{j:}||_{\ell_0}$ to be smaller than $\bar{c}_j \in \mathbb{N}$. Therefore, we can formulate the \textbf{SCONMF} problem as follows:
\begin{align}
    \min_\Theta \min_{W,\hat{H}}\  &D(X,W\hat{H}\Theta) \label{relaxed} 
    \ \textrm{s.t.} \ \hat{H}\hat{H}^{\top}=I_k,  \Theta \in \mathcal{D}_+^n, \\  \nonumber
     &||\hat{H}_{j:}||_{\ell_0} \leq \bar{c}_j,W_{ij},\hat{H}_{ij} \geq 0   .
\end{align}


where in this formulation $H=\hat{H}\Theta$ and $HH^\top \in \mathcal{D}_+^k$. Note that since $\Theta \in \mathcal{D}_+^n,||{H}_{j:}||_{\ell_0} =||\hat{H}_{j:}||_{\ell_0}$ and hence the sparsity constraint can be imposed on $\hat{H}$. In the following theorem, we discuss the uniqueness of the solutions of (\ref{relaxed}). 
\theorem
Let $({W}_1, {H}_1, {\Theta})$ be a solution to the ONMF problem \eqref{relaxed} and ${W}_1{H}_1{\Theta}=\bar{X}$. Then for any other approximation $({W}_2, {H}_2, {\Theta})$ such that ${W}_2 = {W}_1 B^{-1}$ and ${H}_2 = B {H}_1$, it must hold that $B \in \Delta^k_+$. Hence, the factors for any approximation $\bar{X}$ are unique up to generalized permutation.

\begin{proof}
    {\color{black}Suppose $\bar{H}=H_1\Theta,\tilde{H}=H_2\Theta$ and $\bar{H}\bar{H}^\top=\Lambda \in \mathcal{D}_+^k$ where $\Lambda=\text{diag}(\lambda=[\lambda_1,\dots,\lambda_k]),\lambda_i > 0 \ \forall i$. $\tilde{H}=B\bar{H} \Rightarrow \tilde{H}\bar{H}^\top\Lambda^{-1}=B$. Since $B$ is the product of three nonnegative matrices, $B$ itself must be nonnegative as well. On the other hand, $\tilde{H}\tilde{H}^\top \in \mathcal{D}^k_+ \Rightarrow B \bar{H}\bar{H}^\top B^\top = B \Lambda B^\top \in \mathcal{D}^k_+$. Denote $b_i$ to be the $i^\text{th}$ row of $B$. Then, $B \Lambda B^\top \in \mathcal{D}^k_+$ mandates $(\lambda \odot b_i)b_j^\top =0 \ \forall \ i\neq j$. Since $\lambda \odot b_i$ and $b_j$ are nonnegative vectors, this results in each column of $B$ to have exactly one non-zero value. As no $b_i$ is entirely zeros to preserve the rank $k$ of $\tilde{W}$, each column and row of $B$ will have exactly one nonnegative value, implying $B \in \Delta_+^k$. }  
\end{proof}

Now, we show that the ONMF problem (\ref{relaxed}) can be interpreted as a CCFLP. To do so, we first define the FLP problem as follows: given a set of \( k \) facilities \( \{f_j\}_{j=1}^k \) and a set of consumer nodes located at \( \{x_i\}_{i=1}^n \), we aim to assign facilities to consumer nodes via a binary assignment matrix $\Psi=[\psi_{j|i}]\in \{0,1\}^{k\times n}$, where the binary variable $\psi_{j|i}\in \{0,1\}$ is $1$ only when $i^{\text{th}}$ data point is assigned the $j^{\text{th}}$ feature $f_j$, and simultaneously find the optimal facility locations \( \{y_j\}_{j=1}^k \) such that the total distance of nodes to their assigned facilities is minimized. In the CCFLP, also the \emph{density} of each feature, defined as the total number of nodes assigned to it, is set to be upper bounded by predefined capacities $\bar{c}_j$. Hence, the CCFLP is formulated as follows:
\begin{align}
    \min_{\Psi,y_j} \sum_{i=1}^n \bigg\| x_i - \sum_{j=1}^k \psi_{j|i} y_j \bigg\|^2_2 \nonumber \\ \text{s.t.} \ \Psi\Psi^\top \in \mathcal{D}^k_+,  \sum_i \psi_{j|i} \leq \bar{c}_j \ \forall j \label{FLP formulation}
\end{align}
where the constraint $\Psi\Psi^\top \in \mathcal{D}^k_+$ is to ensure each node is assigned to exactly one facility.
\begin{theorem}
    The ONMF problem (\ref{relaxed}) can be interpreted as a CCFLP problem (\ref{FLP formulation}) if $D(\cdot,\cdot)$ is taken to be the squared Frobenius norm $||\cdot||_F^2$. \label{main theorem}
\end{theorem}
\begin{proof}
     Let \( Y = [y_1 ~ \dots ~ y_k] \in \mathbb{R}_+^{d \times k} \) denote the facility location matrix. Then, interpret the columns of the nonnegative matrix \( X \in \mathbb{R}_+^{d \times n} \) as the positions of the consumer nodes \( \{x_i\} \). Define the transformations:
    \begin{align}
W := Y {C}^{1/2}, \quad \hat{H} := {C}^{-1/2} \Psi. \label{trans}
\end{align}
 where \( {C} = \mathrm{diag}({c}_j) \in \mathbb{R}^{k \times k} \), ${c}_j=||\Psi_{j:}||_{\ell_0}= \sum_i \psi_{j|i}$.
 With this transformation, $W,\hat{H}$ are nonnegative, the sparsity constraint $||\hat{H}_{j:}||_{\ell_0} \leq \bar{c}_j$ is satisfied by the equivalent capacity constraint $\sum_i \psi_{j|i} \leq \bar{c}_j$, and also orthogonality $\hat{H}\hat{H}^\top=I_k$ is satisfied. With $\Theta=I_n$ fixed, it follows that:
 \[
     D(X, W\hat{H}\Theta)=\|X - Y\Psi\|_F^2=\sum_{i=1}^n \bigg\| x_i - \sum_{j=1}^k \psi_{j|i} y_j \bigg\|^2_2.
\]
which is identical to the cost of CCFLP (\ref{FLP formulation}).

  Therefore, we can pose the SCONMF as a CCFLP to find $W,\hat{H}$ with fixed $\Theta=I_n$, and then solve the outer minimization in (\ref{relaxed}) by minimizing $E=||X-W\hat{H}\Theta||_F^2$ w.r.t. $\Theta$ which yields:
 \begin{align}
     \Theta = \text{diag}\left(\frac{x_i^{\top} W\hat{H}_{:i}}{||W\hat{H}_{:i}||^2_2}\right) \quad 1\leq i\leq n.
 \end{align}
 where $\hat{H}_{:i}$ denotes the $i^\text{th}$ column of $\hat{H}$.
\end{proof}
\remark
The quantity $c_j=||\hat{H}_{j:}||_{\ell_0}$ measures the \textit{density} of the $j^\text{th}$ feature—that is, how many data points are effectively assigned to it. Imposing upper bounds on ${c}_j$ limits the number of assignments, thereby promoting sparsity in the learned features.

\remark  For the $W$-orthogonal case, the data matrix $X \leftarrow X^{\top}$ is transposed. Following the same reasoning, $W^{\top}W=I$ putting $W \leftarrow ({{C}^{-1/2}\Psi})^{\top}$ and $H \leftarrow ({Y {C}^{1/2}})^{\top}$.

\section{Proposed Solution} \label{solution}
In this section, we propose our solution for the CCFLP and analyze the consequences of adapting it to the ONMF problem as described in theorem \ref{main theorem}.
\subsection{MEP-based Solution With No Capacity Constraints}\label{sec:solution}
The FLP (and therefore ONMF) problems are non-convex, $\mathcal{NP}$-hard optimization problems, where a great deal of complexity arises from the constraint on decision variables $\psi_{j|i}$  to be binary variables. The literature often involves heuristics to address such problems. In our solution, these hard assignments $\psi_{j|i}$ are initially relaxed by soft assignments $p_{j|i}\in [0,1]$, where the probability mass function (PMF) $p_{j|i}$ associates $i^{th}$ data point $x_i$ to the feature $w_j$. These PMFs however, converge to binary values through a process of \emph{annealing} which is the strength of our method (will be explained in this section). According to (\ref{ONMF}), we can reformulate the ONMF as 
\begin{align}\label{Fboth}
&\min_{\{p_{j|i}\},\{w_j\}} \mathcal{F}=\mathcal{D}(\{p_{j|i}\},\{w_j\})-\frac{1}{\beta}\mathcal{H}(\{p_{j|i}\}),\\ \nonumber
&\text{s.t.} \sum_j p_{j|i}=1 \ \forall i, \, \ 0\leq p_{j|i} \leq 1 \ \forall i,j \nonumber
\end{align}
where $\mathcal{D}(\{p_{j|i}\},\{w_j\})=\frac{1}{n}\sum_{i=1}^n\sum_{j=1}^k p_{j|i}d(x_i,w_j)$ is the expected value of the cost function in (\ref{ONMF}), while the Shannon entropy  $\mathcal{H}(\{p_{j|i}\})=-\frac{1}{n}\sum_{i=1}^n \sum_{j=1}^k p_{j|i}\log p_{j|i}$ is a measure of randomness (uncertainty) of the associated PMF $\{p_{j|i}\}$.  $1/\beta$ characterizes the relative importance of the target cost function and the extent of randomness introduced in the formulation due to the PMF.

The optimal solution of (\ref{Fboth}) has been derived in \cite{rose} as follows:
\begin{align}\label{Gibbs}
&p_{j|i}=\frac{ \alpha_je^{-\beta\|x_i-w_j\|^2}}{\sum_{m=1}^k \alpha_me^{-\beta\|x_i-w_m\|^2}} \\
& w_j=\sum_{i=1}^n p_{i|j}x_i, \quad  \alpha_j=\frac{1}{n}\sum_{i=1}^np_{j|i}, \label{optsol}
\end{align}
where $p_{i|j}=p_{j|i}/n\alpha_j$.  Thus, we get optimal (local) solutions for $\{p_{j|i}\}$ and $\{w_j\}$ at a fixed $\beta$, by iterating between equations (\ref{Gibbs},\ref{optsol}) until convergence. For a proof of convergence, see section IV of \cite{1272864}.

In this method, $\beta$ is increased geometrically from a small value upto a maximum value, i.e. $\beta_{t+1} = \beta_t\zeta$, $\zeta>1$, where at each $\beta_{t+1}$ the solution from $\beta_t$ is used as the initial values for $p_{j|i}$ and $w_j$. This process is referred to as \emph{annealing}. See section \ref{sec: phase trans} for a complete analysis.

\remark
When $\beta \rightarrow \infty$, (\ref{Gibbs}) implies that $p_{j|i} \in \{0,1\}$, i.e. relaxed associations become binary at very large $\beta$. In other words, although we initially relax the binary associations $\psi_{j|i}$ to probabilities $p_{j|i}$, the annealing process forces the relaxed associations to converge to binary values, which is remarkably favorable for the purpose of FLP. This results in a mixing matrix $H$ that the nonnegative numbers are always one, as each data point is assigned to exactly one feature. 

\subsection{MEP-Based Solution Under Capacity Constraints} \label{sec:Solution_Ineq}
Building upon the method described in Section \ref{sec:solution}, certain scenarios require additional constraints on the number of data points assigned to the each feature. These constraints can be represented as upper and lower bounds on the weighted \(\ell_0\)-``norm" (capacity) of each feature, denoted as \(\bar{c}_j\).

A notable example of this can be found in the bioinformatics dataset discussed in Section \ref{sec:BioData}, where an effective feature design (i.e., metagene construction) should mitigate extreme imbalances. In particular, each metagene should be assigned a balanced number of genes, preventing scenarios where a few metagenes account for the majority while others contain only a handful. This ensures a more meaningful and interpretable representation of the data.

The MEP-based solution for CCFLP partially resembles that of the unconstrained FLP discussed earlier. However, at each \( \beta \)-iteration, the minimization problem in \eqref{Fboth} is augmented with constraints \( \sum_i p_{j|i} \leq \bar{c}_j \) for all \( j \), making the Gibbs distribution in \eqref{Gibbs} potentially infeasible. These added constraints increase the problem's complexity, as no closed-form solution exists in general. Existing methods—such as penalty-based approaches~\cite{srivastava2022inequality} or SLSQP~\cite{boggs1995sequential}—either fail to enforce constraints or scale poorly with problem size. To address this, we propose a control-theoretic approach inspired by CBFs~\cite{7782377}, assigning the following control dynamics to the decision variables:
\begin{equation}
\begin{aligned}
    &\dot{p}_{j|i} = v_{ij}, \quad   && p_{j|i}(0) = p_{j|i}^0 \in (0,1),  && \forall i, j,\\
    &\dot{w}_j = u_j, \quad && w_j(0) = w_j^0, && \forall j.
\end{aligned}
\label{eq:control_dynamics}
\end{equation}
We now present the following theorem:
\begin{theorem} \label{thrm:main_CCFLP}
Let \( v_{ij}^*(\{p_{j|i}\}, \{w_j\}) \) and \( u_j^*(\{p_{j|i}\}, \{w_j\}) \) denote the solution to the following quadratic program, defined for any feasible \( \{p_{j|i}\} \) and \( \{w_j\} \):
\begin{subequations} \label{eq:QP}
\begin{align}
    \min_{\{v_{ij}\}, \{u_j\}} & \quad \sum_{i,j} \|v_{ij}\|^2 + \sum_j \|u_j\|^2 + q \, \delta^2 \label{eq:QP_obj}\\
    \text{s.t.} \quad 
    &\dot{V}(\{p_{j|i}\}, \{w_j\}) \leq -\gamma V(\{p_{j|i}\}, \{w_j\}) + \delta \label{eq:descent_V}\\
    &\dot{\phi}(\{p_{j|i}\}_j) = 0, \quad \forall i, \label{eq:phi} \\
    &\dot{\psi}(\{p_{j|i}\}_i) \geq -\lambda \, \psi(\{p_{j|i}\}_i) \quad \forall j, \\
    &\dot{\xi}(p_{j|i}) \geq -\mu \, \xi(p_{j|i}), \quad \forall i, j,
\end{align}
\end{subequations}
where \( \gamma, \lambda, \mu, q > 0 \) are design constants, and the functions are defined as:
\begin{align}
    &V(\{p_{j|i}\}, \{w_j\}) = \mathcal{F}(\{p_{j|i}\}, \{w_j\}) + \frac{\log k}{\beta}, \label{def:Lyapunov} \\
    &\phi(\{p_{j|i}\}_j) = \sum_j p_{j|i} - 1, \\
    &\psi(\{p_{j|i}\}_i) = c_j - \sum_i p_{j|i}, \\
    &\xi(p_{j|i}) = p_{j|i}(1 - p_{j|i}).
\end{align}
Then, if the initial conditions \( \{p_{j|i}^0\} \) satisfy
\[
\begin{cases}
\phi(\{p_{j|i}^0\}_j) = 0 & \forall i, \\
\psi(\{p_{j|i}^0\}_i) \geq 0 & \forall j, \\
\xi(p_{j|i}^0) > 0 & \forall i, j,
\end{cases}
\]
the trajectories \( \{p_{j|i}(t)\} \) and \( \{w_j(t)\} \) of \eqref{eq:control_dynamics} generated under this control law converge to a KKT point of the CCFLP (\ref{FLP formulation}).
\end{theorem}

\begin{proof}
A verbal explanation is provided here; a rigorous proof of convergence for the general constrained optimization setting—of which this problem is a special case—is given in the concurrent work \cite{bayati2025controlbarrierfunctionapproach}.

Note that \( V \) is a shifted version of the free energy \( \mathcal{F} \) to ensure non-negativity for all \( \{p_{j|i}\} \) and \( \{w_j\} \). It serves as a CLF-like function, with constraint~\eqref{eq:descent_V} promoting descent of \( \mathcal{F} \) wherever feasible. Importantly, \( V \) is radially unbounded with respect to each feature \( w_j \), i.e., \( V \to \infty \) as \( \|w_j\| \to \infty \). The function \( \phi \) encodes the equality constraints, and constraint~\eqref{eq:phi} ensures it remains satisfied for all \( t \geq 0 \). Finally, \( \psi \) and \( \xi \) act as CBFs to enforce capacity constraints along the trajectory and maintain \( p_{j|i}(t) \in (0,1) \) for all \( i, j \) and \( t \geq 0 \).

If \( \{p_{j|i}\} \) and \( \{w_j\} \) do not correspond to a KKT point of the CCFLP, then there exists a nonzero descent direction \( \{\tilde{v}_{ij}\} \), \( \{\tilde{u}_j\} \) that preserves both equality and inequality constraints (See ~\cite{bayati2025controlbarrierfunctionapproach}). A sufficiently small step in this direction yields a strictly lower cost in~\eqref{eq:QP_obj} than the trivially feasible zero control. Thus, the optimal controls \( v_{ij}^*(\{p_{j|i}\}, \{w_j\}) \), \( u_j^*(\{p_{j|i}\}, \{w_j\}) \) must satisfy \( \dot{V}(\{p_{j|i}\}, \{w_j\}) < 0 \). Moreover, these controls never yield \( \dot{V} > 0 \), so \( \dot{V} \leq 0 \) for all \( \{p_{j|i}\} \), \( \{w_j\} \), with strict inequality away from KKT points.

By Theorem~1 of~\cite{6760327}, this control law is locally Lipschitz continuous. As \( V \) is radially unbounded in \( \{w_j\} \), trajectories are well-defined and bounded for all \( t \geq 0 \)~\cite{khalil2002nonlinear}, and by LaSalle’s Invariance Principle~\cite{khalil2002nonlinear}, the system converges to the KKT points of the CCFLP.
\end{proof}

\remark
KKT points of the CCFLP correspond to assignments \( \{p_{j|i}\} \) that respect feature capacity constraints, while the features \( \{w_j\} \) still satisfy the weighted centroid condition given in \eqref{optsol}.


When capacity constraints are enforced, even when \(\beta \to \infty\), the nearest feature to a given \(x_i\) may lack sufficient capacity to fully accommodate it. In such cases, only a fraction of \(x_i\) is assigned to the closest feature until its capacity is exhausted; any remaining portion is then allocated to the next-closest feature with available capacity. Consequently, the resulting probability distributions may not converge to binary vectors for all data points. To address this issue, one can either relax the capacity constraints by replacing them with soft constraints—introducing a slack variable and penalizing its violation in the cost function—or, alternatively, post-process the resulting distributions by projecting them onto the nearest binary vector.


The pseudocode for the control-theoretic approach to solving the CCFLP is provided below.

\begin{algorithm} \label{Alg:CCFLP}
\caption{Control-theoretic Approach for CCFLP}
\begin{algorithmic}[1]
\State \textbf{Input:} Initial conditions \( \mathbf{p}_0 := \{p_{j|i}^0\} \), \( \mathbf{w}_0 := \{w_j^0\} \)
\State \textbf{Parameters:} \( \gamma, \lambda, \mu, q, \beta_0, \beta_{\max}, \alpha > 1, dt \)
\State \textbf{Initialize:} \( \mathbf{p} \gets \mathbf{p}_0 \), \( \mathbf{w} \gets \mathbf{w}_0 \), \( \beta \gets \beta_0 \)
\While{\( \beta \leq \beta_{\max} \)}
    \While{\( \mathbf{p} \) and \( \mathbf{w} \) have not converged}
        \State Compute optimal controls as in QP \eqref{eq:QP}:
        \[
        \mathbf{v} := \{v_{ij}^*(\mathbf{p}, \mathbf{w})\}, \quad \mathbf{u} := \{u_j^*(\mathbf{p}, \mathbf{w})\}
        \]
        \State Update states:
        \[
        \mathbf{p} \gets \mathbf{p} + \mathbf{v} \cdot dt, \quad \mathbf{w} \gets \mathbf{w} + \mathbf{u} \cdot dt
        \]
    \EndWhile
    \State Increase \(\beta\): \( \beta \gets \alpha \cdot \beta \)
\EndWhile
\State \textbf{Output:} \( \mathbf{p} \) and \( \mathbf{w} \) satisfying KKT conditions for \(\beta_{\max}\)
\end{algorithmic}
\end{algorithm}

To enhance computational efficiency, the step size \( dt \) is adapted dynamically based on the current values of \( \mathbf{p}, \mathbf{w}, \mathbf{u}, \mathbf{v} \), starting from an initial step size \( dt^0 \). This strategy follows the principle of adaptive step sizing commonly used in gradient-based optimization. In our simulations presented in Section~\ref{sec:BioData}, we employ the approach introduced in~\cite{pmlr-v119-malitsky20a} to update the step size.

\subsection{Phase Transitions and True Number of Features} \label{sec: phase trans}
 The proposed method in section \ref{solution} evolves hierarchically with respect to $\beta$, as $\beta$ is increased from small ($\approx 0$) to a high ($\approx \infty$) value; at each iteration the solutions from the previous iteration are used for initialization. Note that at initial iterations (when $\beta\approx 0 $), higher emphasis is given to the randomness of associations (characterized by $\mathcal{H}$ in (\ref{Fboth})); hence the ensuing solutions in the optimally-weighted case are uniform PMF $p_{j|i}\approx \frac{1}{k} \; \forall i$, and all $w_j$ are coincident the centroid of data points $x_i$. This is evident by looking at equations (\ref{Gibbs}) and (\ref{optsol}) at $\beta\approx 0$. Therefore all the $k$ features are coincident at the centroid when $\beta\approx 0$. This can be also explained since the term  $\sum_j e^{-\beta \|x_i-w_j\|^2}$ in ${\mathcal{F}}$ cannot distinguish different summands since $\|x_i-w_j\|^2\ll \frac{1}{\beta}$ for all $j$ at each $i$. Thus $1/\beta$ acts as a resolution measure on reconstruction error (cost value in (\ref{ONMF})); and when this resolution yardstick is too large, one feature is enough to {\em achieve} that resolution in reconstruction error. As $\beta$ is increased, this resolution yardstick becomes smaller (finer);  whereby  $\|x_i-w_j\|^2$ for different $j$s are more {\em distinguishable}. As $\beta$ is increased from $0$, there is a critical value $\beta_{cr}$ beyond which it is not possible to achieve the now smaller resolution on reconstruction error by a {\em single} distinct value of $w_j$ but requires at least two distinct features (this can be used as a phase transition condition). Thus as the resolution ($1/\beta)$ or reconstruction error bound is decreased; more number of {\em distinct} features appear in the optimal solutions at successive values of $\beta_{cr}$.
 \par
In the context of NMF and ONMF algorithms, it is common to assume that the number of features is known a priori, or to  constrain it in some way. However, we can utilize the phase  transition concept to identify the true number of features present in a dataset. We adapt the notions developed in \cite{truenumber} in the context of the clustering problem to our problem.   Based on the phase transitions at successive critical temperatures, we define a measure  $\Delta(m)={\beta_{cr}(m+1)}/ {\beta_{cr}(m)}$ that quantifies {\em persistence} of  $m$ distinct features -  Here $\Delta(m)$ quantifies the range of reconstruction error bounds (characterized by $1/\beta$) for which $m$ is the smallest number of distinct features  necessary (and enough) to guarantee  those bounds.  The true number of features is then defines as one that persists for the largest range of reconstruction errors. More precisely, {\em true} number $m^\star$ of features is one that satisfies $m^\star=\mbox{arg}\max \Delta(m)$.
\section{Evaluation Setup} \label{sec:setup}
To evaluate our algorithm and compare it with other existing algorithms\footnote{The dataset and an implementation of the algorithm are available on a Github repository at :\url{https://github.com/salar96/MEP-Orthogonal-NMF}.}, we use the following four metrics:\\ (a) {\it Reconstruction error}: given by $E={||X-WH||_F}/{||X||_F}$.
(b) {\it Orthogonality}: 
calculated as $O=1-{||HH^{\top}-HH^{\top} \odot I||_F}/{||HH^{\top}||_F}$.
(c) {\it Sparsity}: defined as $S=1-\frac{1}{kn}\sum_{j=1}^k ||H_{j:}||_{\ell_0}$.
(d) {\it Execution time ($T$)}: The total time elapsed in seconds\footnote{All of the algorithms are executed using an Intel\textsuperscript{\textregistered} Core\textsuperscript{\texttrademark} i7-4790 CPU (@ 3.60 GHz) and each is run 5 times. The reported values for all the metrics are the average values over all runs.}.
\par We have chosen similar algorithms in the literature to compare our algorithm with. These include methods in \cite{app} (ONMF-apx), \cite{HALS} (HALS), \cite{nhl} (NLHN), \cite{onmfa} (ONMF-A), \cite{ding} (ONMF-Ding), \cite{emonmf} (ONMF-EM), and \cite{pnmf} (PNMF). We call our method \textbf{MEP-ONMF}.
The evaluation is done in two scenarios. In scenario one, we ran the algorithms on random synthetic matrices; Specifically, the columns of the data matrix $X$ were sampled from a Gamma distribution, with the probability density function defined as $P(x)=x^{\alpha-1}{e^{\frac{-x}{\theta}}}/{\theta^{\alpha}\Gamma(\alpha)}$ where $\Gamma(\cdot)$ is the Gamma function. Here we have chosen $\alpha=10$ and $\theta=1$. Additionally, a uniform random noise was incorporated into the matrix to further increase the diversity of the data. In this scenario, we use $k=20$ as the inner dimension (rank of the factors) in all of the datasets. Hence, four randomly generated $d\times n$ datasets are generated, where the ($d,n$) values for datasets 1-4 are (10,1000), (20,2000), (50,4000), and (100,10000) respectively. In the second scenario, we utilized a standard bioinformatics dataset \cite{Baranzini2004} (dataset 5), which contains microarray data collected from patients over different time periods. Microarrays represent gene expression levels as nonnegative numerical values, providing insights into how genes are expressed under various conditions. These data are typically structured in gene-sample or gene-time matrices \cite{Esposito_2021}, where rows correspond to genes and columns represent samples or time points.

State-of-the-art approaches for analyzing such datasets often employ ONMF techniques, aiming to decompose the original matrix into two factor matrices: one representing a set of ``metagenes" (features) and the other quantifying their contributions across samples or time points. These metagenes are linear combinations of the original genes and collectively describe the entire microarray dataset. A key characteristic of orthogonal NMF is that it enforces non-overlapping features, meaning each gene is assigned to only one metagene. This property enhances interpretability by ensuring that the extracted metagenes are distinct and biologically meaningful.

\section{Results and Discussion}
\subsection{Synthetic data}

\begin{table*}[!t] 
\centering
\caption{Simulation results for different datasets. For definition of metrics, refer to section \ref{sec:setup}.}
\label{table:results_syn}
\resizebox{\textwidth}{!}{%
\begin{tabular}{l|ccccc|ccccc|ccccc|ccccc}
\toprule
\multirow{2}{*}{} & \multicolumn{5}{c|}{\textbf{E(\%)}} & \multicolumn{5}{c|}{\textbf{O(\%)}} & \multicolumn{5}{c|}{\textbf{S(\%)}} & \multicolumn{5}{c}{\textbf{T(s)}} \\ 
\cmidrule(lr){2-6} \cmidrule(lr){7-11} \cmidrule(lr){12-16} \cmidrule(lr){17-21}
 Dataset & \textbf{1} & \textbf{2} & \textbf{3} & \textbf{4} & \textbf{5} & \textbf{1} & \textbf{2} & \textbf{3} & \textbf{4} & \textbf{5} & \textbf{1} & \textbf{2} & \textbf{3} & \textbf{4} & \textbf{5} & \textbf{1} & \textbf{2} & \textbf{3} & \textbf{4} & \textbf{5} \\
\midrule
\textbf{MEP\_ONMF} & 0.026 & \textbf{0.027} & \textbf{0.028} & \textbf{0.028} & \textbf{30.678} & \textbf{100} & \textbf{100} & \textbf{100} & \textbf{100} & \textbf{100} & \textbf{95} & \textbf{95} & \textbf{95} & \textbf{95} & \textbf{66} & \textbf{0.040} & \textbf{0.085} & \textbf{0.252} & \textbf{0.965} & 0.033 \\
ONMF\_apx \cite{app} & 0.026 & \textbf{0.027} & \textbf{0.028} & \textbf{0.028} & 30.707 & \textbf{100} & \textbf{100} & \textbf{100} & \textbf{100} & \textbf{100} & \textbf{95} & \textbf{95} & \textbf{95} & \textbf{95} & \textbf{66} & 0.133 & 0.264 & 0.420 & 1.239 & 0.018 \\
ONMF\_Ding \cite{ding} & 2.520 & 0.107 & 3.092 & 5.375 & 29.651 & 85 & 97 & 90 & 87 & 77 & 78 & 78 & 86 & 87 & 36 & 19.193 & 46.584 & $\approx$300 & $\approx$1400 & 0.311 \\
ONMF\_A \cite{onmfa} & 4.235 & 4.846 & 2.130 & 4.062 & 30.006 & 74 & 81 & 90 & 88 & 80 & 66 & 66 & 61 & 69 & 30 & 1.789 & 5.930 & 17.181 & 93.992 & 0.054 \\
PNMF \cite{pnmf} & 4.305 & 4.975 & 7.345 & 6.341 & 30.250 & 81 & 86 & 82 & 90 & 84 & 79 & 82 & 82 & 87 & 42 & 31.185 & $\approx$130 & $\approx$530 & $\approx$3600 & 0.325 \\
NHL \cite{nhl} & 5.166 & 6.324 & 6.538 & 7.718 & 30.432 & 82 & 84 & 88 & 86 & 76 & 80 & 85 & 86 & 89 & 33 & 27.051 & $\approx$190 & $\approx$550 & $\approx$3560 & 0.195 \\
ONMF\_EM \cite{emonmf} & 9.522 & 8.782 & 9.839 & 9.491 & 31.169 & \textbf{100} & \textbf{100} & \textbf{100} & \textbf{100} & \textbf{100} & \textbf{95} & \textbf{95} & \textbf{95} & \textbf{95} & \textbf{66} & 0.086 & 0.201 & 0.666 & 3.133 & 0.012 \\
HALS \cite{HALS} & \textbf{0.006} & 0.483 & 1.042 & 0.049 & 32.663 & 13 & 13 & 26 & 46 & \textbf{100} & 24 & 29 & 52 & 56 & \textbf{66} & 2.099 & 2.783 & 26.565 & 90.312 & 0.019 \\
iONMF \cite{IONMF} & 1.917 & 5.006 & 7.098 & 8.947 & 28.723 & 10 & 11 & 14 & 17 & 34 & 0 & 0 & 0 & 0 & 3 & 0.071 & 0.139 & 0.314 & 1.269 & \textbf{0.007} \\
NMF \cite{Lee_1999} & 0.028 & 0.091 & 0.210 & 0.528 & 28.229 & 10 & 14 & 28 & 46 & 30 & 0 & 0 & 16 & 37 & 4 & 0.348 & 4.788 & 30.208 & $\approx$300 & 0.322 \\
\bottomrule
\end{tabular}%
}
\end{table*}
The results for synthetic matrices are shown in Table \ref{table:results_syn}. The best values in each column are bolded. For the dataset one, the underlying data is relatively low dimensional. Our proposed method demonstrates the fastest performance time (0.04s vs. average 9.11s), as well as the highest levels of orthogonality and sparsity when compared to other methods. The HALS method exhibits a lower reconstruction error, but at the cost of compromised orthogonality and sparsity (13\% and 24\% respectively). Our method not only guarantees orthogonality, but also yields a smaller reconstruction error in comparison to the original NMF method ($\approx$7\%) which does not have a guarantee of orthogonality. This pattern is also observed in datasets two and three, where the dimensions of the datasets have increased significantly. In these cases, our method achieves the lowest reconstruction error among all methods (average 0.03 vs. average 3.78). The ONMF-EM also yields orthogonality and high sparsity, however, it results in a higher reconstruction error (average 9.31). Finally, in the dataset four, data dimensions are significantly large, and our method is demonstrated to scale well ($\approx$1s), as the performance time remains efficient in comparison to ONMF-ding, PNMF, and NHL. Our method achieves the best reconstruction error (0.03 vs. average 4.73) with full orthogonality, the highest sparsity, and the fastest run-time.\par
\subsection{Standard Bioinformatics Dataset} \label{sec:BioData}
\begin{figure}[t]
    \centering
    \includegraphics[trim={0.3cm 5cm 0.3cm 6cm},clip,width=1\columnwidth]{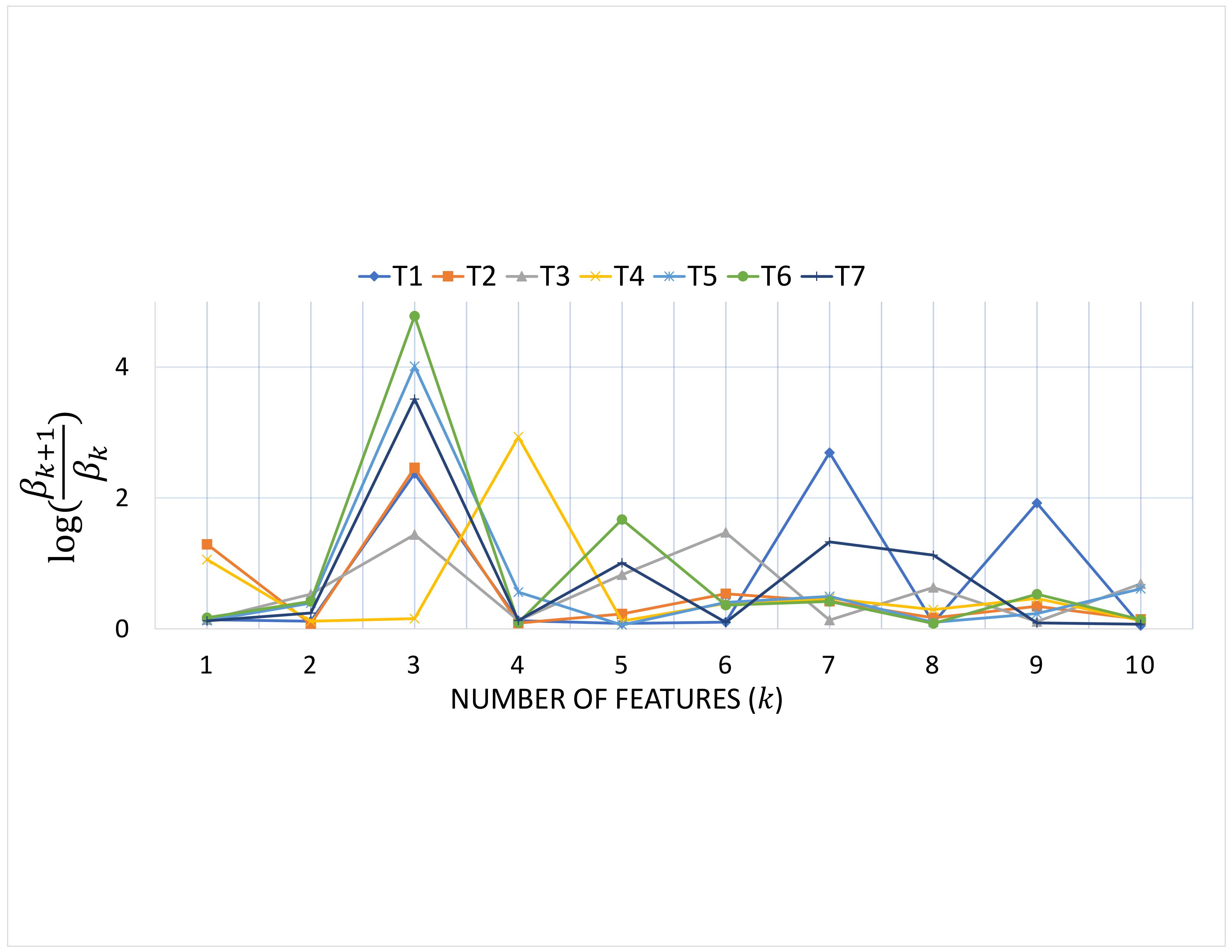}
    \caption{The logarithm of the fraction between successive critical $\beta$s over all time periods. The $k^{th}$ value on the x-axis represents the number of features, while The y-axis shows log($\frac{\beta_{k+1}}{\beta_k}$) where $\beta_k$ is the critical $\beta$ value at which the $k^{th}$ feature split happens.}
    \label{fig:criticals}
\end{figure}

In this scenario, we have used MEP-ONMF to extract the main features (i.e. metagenes) and compare our results with other methods. The first step was to determine the number of metagenes that we want to calculate. As explained in section \ref{sec: phase trans}, the methodology of MEP-ONMF provides a feasible way to determine the true number of features in a dataset. Simply by looking at the critical $\beta$s diagram ($\beta$s at which a feature split has happened), we can determine the true number of features, and that is when a large gap is seen between two consecutive values. The logarithmic difference between successive critical $\beta$ values for all time periods is depicted in Fig. \ref{fig:criticals}. If a significant spike is observed at the $k^{th}$ split, indicating a transition from $k$ features to $k+1$ features, it can be concluded that selecting $k$ features yields the most persistent factorization and can thus be considered as the true number of features in the dataset. In almost all time periods, there is a large spike at the third split, except for one ($T_4$) where this gap happens at the fourth. Therefore, we can conclude that the true number of metagenes in this dataset is 3, which approves the number used in previous works.
Tabel \ref{table:results_syn} shows the results of the simulation on the dataset 5, averaged over all the 7 time period and 5 runs for each algorithm in total.


The results of this table indicate that our proposed algorithm is able to generate perfectly orthogonal metagenes with higher sparsity compared to other methods (66\% vs. average 38\%). Furthermore, the algorithm results in a smaller reconstruction error ($\approx$1.5\% on average) but slightly slower performance time  compared to other methods that do not enforce the constraint of orthogonality. 
\par
\begin{figure}[t]
    \centering \includegraphics[width=1\columnwidth]{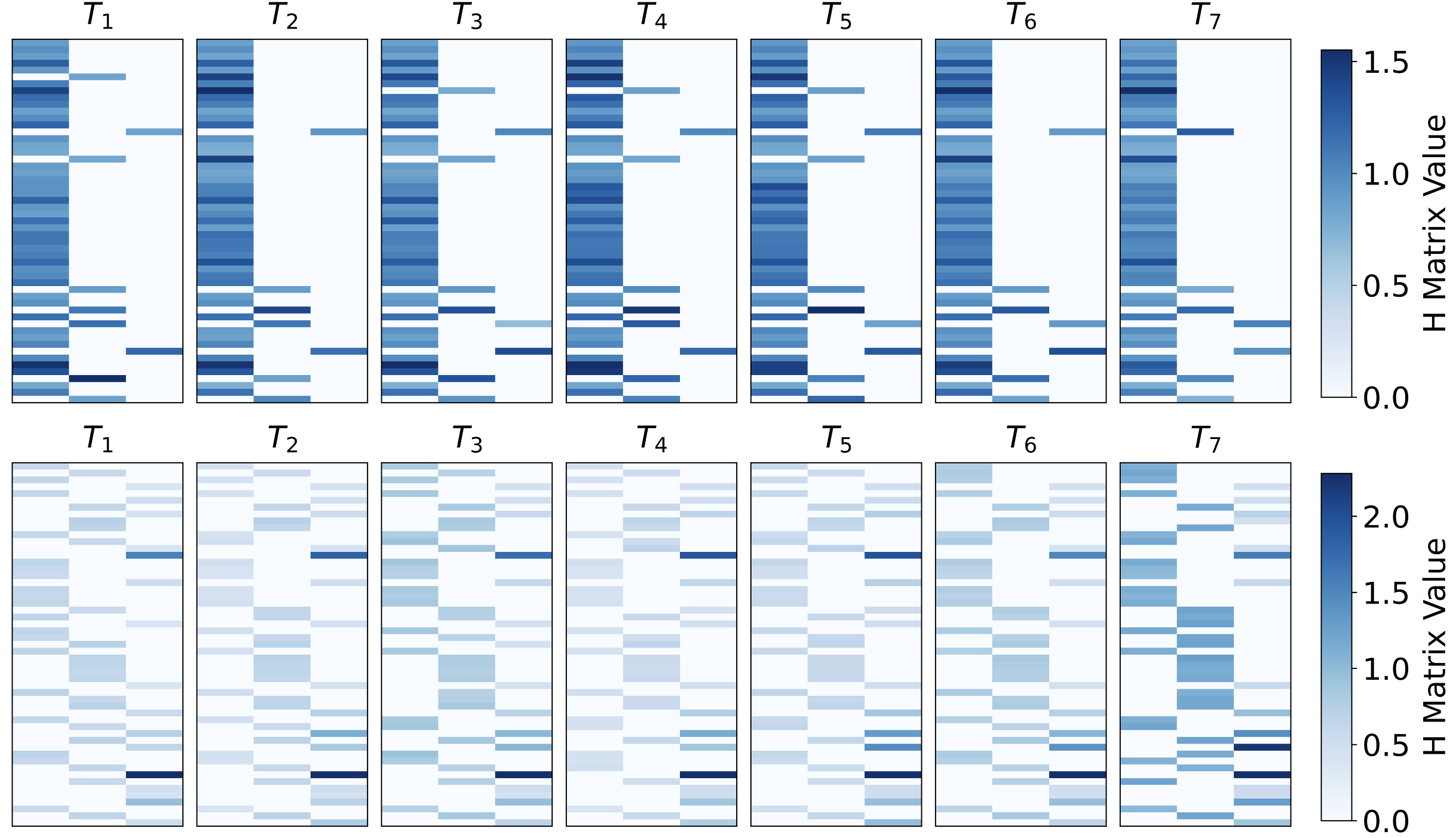}
    \caption{Orthogonal features extracted using MEP-ONMF across seven time periods of the standard bioinformatics dataset. The top row represents the unconstrained case, while the bottom row corresponds to the constrained setting (\(\bar{c}_j/n \leq 0.35\) for all \(j\)). The y-axis denotes genes, and the x-axis represents metagenes.
    }
    \label{fig:metas}
\end{figure}

 %
Figure \ref{fig:metas} illustrates the metagenes computed using MEP-ONMF in both unconstrained (upper plot) and inequality-constrained (lower plot) settings, as described in Sections \ref{sec:solution} and \ref{sec:Solution_Ineq}, across seven time periods. In the constrained case, the \(\ell_0\)-``norm" of each metagene is limited to 0.35$n$ (i.e., \(\bar{c}_1, \bar{c}_2, \bar{c}_3 \leq 0.35n\)). The corresponding cost values and maximum \(\ell_0\)-norms of the metagenes for both cases are summarized in Table \ref{tab:comparison}.
\begin{table}[ht]
    \centering
    \caption{ONMF on a Standard Bioinformatics Dataset: Unconstrained vs. Capacity-Constrained Cases}
    \textbf{Unconstrained Case}
    \vspace{0.2cm}
    
    \begin{tabular}{|c|c|c|c|c|c|c|c|}
        \hline
        & \(T_1\) & \(T_2\) & \(T_3\) & \(T_4\) & \(T_5\) & \(T_6\) & \(T_7\) \\
        \hline
        \(\mathcal{D}\) & 13.8  & 12.2 & 22.0 & 16.4 & 23.5 & 19.9 & 17.1  \\
        \hline
        \(\max_j {c}_j/n\) & 0.83  & 0.87 & 0.83 & 0.83 & 0.83 & 0.87 & 0.87 \\
        \hline
    \end{tabular}
    
    \vspace{0.4cm} 

    \textbf{Inequality-Constrained Case (\(c_j/n \leq 0.35\))}
    \vspace{0.1cm}
    
    \begin{tabular}{|c|c|c|c|c|c|c|c|}
        \hline
        & \(T_1\) & \(T_2\) & \(T_3\) & \(T_4\) & \(T_5\) & \(T_6\) & \(T_7\) \\
        \hline
        \(\mathcal{D}\)   & 47.0 & 56.7 & 50.6 & 64.7 & 64.5 & 48.8 & 48.2 \\
        \hline
        \(\max_j {c}_j/n\) & 0.36 & 0.36 & 0.36 & 0.36 & 0.36 & 0.36 & 0.36 \\
        \hline
    \end{tabular}
    \vspace{0.2cm}

    \label{tab:comparison}
\end{table}

Note that the cost values presented in Table \ref{tab:comparison} were computed after normalizing the data to the range \([1,10]\) using the transformation
\[
X' = \frac{X - X_{\min}}{X_{\max} - X_{\min}}(10-1) + 1,
\]
where \(X_{\min}\) and \(X_{\max}\) denote the minimum and maximum values of the original dataset, respectively. Furthermore, the slight capacity constraint violation (0.36 instead of 0.35) is due to projecting the obtained probability vectors onto the nearest binary vector, as described in Section \ref{sec:Solution_Ineq}.

\section{Conclusion}
This paper introduces a novel framework for solving the SCONMF problem by reformulating it as a CCFLP. The proposed method integrates CBF techniques with a MEP-based facility location framework to jointly enforce non-negativity, orthogonality, and sparsity constraints. Additionally, we present a principled approach to estimate the true rank of the factorization, corresponding to the number of inherent features, which is crucial when this number is unknown. Experimental results demonstrate substantially improved factorizations with significantly lower reconstruction errors while strictly satisfying all imposed constraints.

\section*{Acknowledgment}
The authors would like to thank Moses Charikar and Lunjia Hu for sharing their code in \cite{app}.
\bibliographystyle{IEEEtran}
\bibliography{references}

\end{document}